# Sources of HIV infections among MSM with a migration background: a viral phylogenetic case study in Amsterdam, the Netherlands

Alexandra Blenkinsop[1*], Nikos Pantazis[2], Evangelia Georgia Kostaki[2],
Lysandros Sofocleous[1], Ard van Sighem[3], Daniela Bezemer[3], Thijs van de Laar[4],
Marc van der Valk[3,7], Peter Reiss[5,6], Godelieve de Bree[5,7], Oliver Ratmann[2]
on behalf of the HIV Transmission Elimination AMsterdam Initiative

[1]Department of Mathematics, Imperial College London, London, SW7 2AZ, UK
[2]Department of Hygiene, Epidemiology and Medical Statistics, Medical School, National and Kapodistrian University of Athens, Athens, 115 27, Greece
[3]Stichting HIV Monitoring, Amsterdam, 1105 BD, the Netherlands
[4]Department of Donor Medicine Research, Sanquin, Amsterdam, 1066 CX, the Netherlands
[5]Amsterdam Institute for Global Health and Development, Amsterdam, 1105 BP, the Netherlands
[6]Department of Global Health, Amsterdam University Medical Centers, University of Amsterdam, Amsterdam, 1105 AZ, the Netherlands
[7]Amsterdam Institute for Infection and Immunity, Amsterdam University Medical Centers, Amsterdam, 1105 AZ, the Netherlands
*a.blenkinsop@imperial.ac.uk

## Abstract

**Background** Men and women with a migration background comprise an increasing proportion of incident HIV cases across Western Europe. Several studies indicate a substantial proportion acquire HIV post-migration.

**Methods** We used partial HIV consensus sequences with linked demographic and clinical data from the opt-out ATHENA cohort of people with HIV in the Netherlands to quantify population-level sources of transmission to Dutch-born and foreign-born Amsterdam men who have sex with men (MSM) between 2010-2021. We identified phylogenetically and epidemiologically possible transmission pairs in local transmission chains and interpreted these in the context of estimated infection dates, quantifying transmission dynamics between sub-populations by world region of birth.

**Results** We estimate the majority of Amsterdam MSM who acquired their infection locally had a Dutch-born Amsterdam MSM source (56% [53-58%]). Dutch-born MSM were the predominant source population of infections among almost all foreign-born Amsterdam MSM sub-populations. Stratifying by two-year intervals indicated shifts in transmission dynamics, with a majority of infections originating from foreign-born MSM since 2018, although uncertainty ranges remained wide.

**Conclusions** In the context of declining HIV incidence among Amsterdam MSM, our data suggest whilst native-born MSM have predominantly driven transmissions in 2010-2021, the contribution from foreign-born MSM living in Amsterdam is increasing.

# 1 Background

Incidence of HIV among MSM has declined across Western Europe over the past five years, particularly since the adoption of immediate initiation of combination anti-retroviral therapy (cART) following diagnosis,[1] widespread uptake of pre-exposure prophylaxis (PrEP) as prevention,[2,3] intensified testing programs,[4,5] and further evidence-based prevention strategies.[6] However, incidence remains substantially higher among MSM with a migration background (MSM-MB) compared with native-born residents, suggesting gaps in prevention services.[7] In the aMASE study spanning nine European countries in 2013-2015[8] and further independent studies,[9,10] the majority of MSM with a migration background living with HIV in Europe were estimated to have acquired their infection post-migration. These data prompted us to investigate which population groups are sustaining transmission to MSM with a migration background in their current place of residence. We focus on transmission occurring during 2010-2021 in Amsterdam, the capital of the Netherlands and one of the early epicenters of the HIV epidemic in Europe.

In 2014, the HIV Transmission Elimination Amsterdam initiative (H-TEAM) was established to develop an evidence-led response with the aim of eliminating HIV transmission within the city (`www.hteam.nl`). Following the implementation of several successful interventions,[11,12] Amsterdam has surpassed UNAIDS targets, with 97% of individuals knowing their status, 95% of diagnosed individuals on treatment, and 96% of treated individuals virally suppressed.[13] However, since 2010 over half of new diagnoses among MSM were in men born outside of the Netherlands, despite first-generation migrants comprising only a third of the general population in Amsterdam.

To characterize transmission sources, several studies[14–18] used subtype data of HIV positive individuals as an indicator in Europe since non-B subtypes typically have origins from world regions with higher HIV prevalence, and identified growing proportions of non-B subtypes in particular among MSM,[16,17] suggesting that transmission dynamics may be changing and increasingly originating from source populations with non-B subtypes. It is unclear if these trends reflect a constant rate of external importations in the midst of declining local transmission, an increasing rate of importations, or increased local transmission from MSM with non-B subtypes. Phylodynamic analyses can help resolve these alternative hypotheses, especially when comprehensive collections of background sequences are used to disentangle local, growing transmission chains from importations.[9] A recent phylodynamic analysis from the UK[19] suggested the majority of non-White MSM living with HIV in the United Kingdom acquired their infection from a transmission source of White ethnicity. However estimates were based on subtype B sequences only, and it remains unclear if these findings generalise to all MSM.

Here, we applied pathogen phylodynamic analyses across all predominant subtypes to estimate sources of HIV infection among Dutch-born and foreign-born Amsterdam MSM in 2010-2021. We leveraged infection date estimates, which have been shown to be instrumental for improving source attribution,[20] and then quantified differences in transmission flows between Dutch-born MSM and the primary migrant groups among MSM. Independently, we also investigated subtype frequency trends to corroborate our phylodynamic analyses. With similar trends in incidence observed in migrant MSM across France, the UK and other European countries,[21–23] this study may provide useful insights beyond Amsterdam.

# 2 Methods

## 2.1 The ATHENA cohort study

Data were collected as part of the open, opt-out ATHENA observational cohort study of people living with HIV in the Netherlands, with the earliest date of diagnosis on 12/06/1981.[24] People entering HIV care receive written material about participation in the ATHENA cohort, after which they can consent verbally or elect to opt-out. 2% of eligible participants opted out and 5.2% were lost to follow-up. A database freeze on 01/02/2022 comprised pathogen genomic, epidemiological and clinical data for MSM ever resident in Amsterdam, located using postal codes (PC4) of participants' address at time of registration into the cohort or registration update. MSM were grouped into geographic regions of birth corresponding to the primary migrant groups among MSM living with HIV in Amsterdam: Western Europe, North America & Oceania, Eastern and Central Europe, Suriname & the Dutch Caribbean, South America & the non-Dutch Caribbean, the Middle East & North Africa (MENA), or other world regions (Supplementary Figure S1).

## 2.2 Estimating infection dates and HIV incidence

Our study population comprised Amsterdam MSM estimated to have acquired HIV between January 1 2010 to December 31 2021. We estimated dates of infection for each diagnosed Amsterdam MSM from longitudinal viral loads and CD4 counts of participants,[9,25] and incidence estimates were derived after accounting for estimated probabilities of remaining undiagnosed by database freeze. Annual HIV prevalence was estimated by summing incidence and subtracting individuals in the cohort known to have died (Supplementary Material Section S1).

## 2.3 Phylogenetic analysis

Partial HIV-1 polymerase (pol) sequences could be obtained for 40% of participants in the Netherlands, and 55% of participants with an Amsterdam postal code. Over 80,000 international *pol* sequences were obtained from the Los Alamos database (www.hiv.lanl.gov) for phylogenetic background, with the closest sequences to the ATHENA sequences selected using *BLAST* (v2.10.0).[26] ATHENA sequences and background sequences were aligned with *Virulign*[27] and MAFFT[28] with manual curation, and subtyped using *COMET* (v2.3)[29] with unassigned subtypes verified with *REGA* (v3.0).[30] Phylogenetic trees were inferred using *FastTree* (v2.1.8)[31] for all major subtypes and circulating recombinant forms (CRFs) in Amsterdam (B, 01AE, 02AG, A1, C, G, D, 06cpx). Other subtypes were excluded from subsequent analyses due to small numbers of sequences. Ancestral state reconstruction was carried out with *phyloscanner* (v.1.8.0).[32] Amsterdam MSM transmission chains were identified as groups of tips connected by internal nodes assigned the same ancestral host state[9]

## 2.4 Estimating subtype frequencies

To estimate annual subtype frequencies as a crude indicator of transmission sources, we used as denominator the estimated number of incident cases per year including estimated undiagnosed infections by database freeze, and as numerator the number of sequenced incident cases with B or non-B subtypes plus estimated numbers of undiagnosed/unsequenced incident cases of the same subtype. We estimated the latter assuming subtype distributions were as seen among diagnosed and sequenced cases; or assuming conservatively for our hypotheses that all undiagnosed incident cases were of subtype B (Supplementary Material Section S2).

## 2.5 Identifying phylogenetically possible transmission pairs

From the phylogenies we extracted all phylogenetically possible transmission pairs in the same Amsterdam transmission chain for incident cases since 2010, such that the estimated infection date of the source preceded that of the incident case. Pairs were excluded if participant metadata indicated the source had died or not yet migrated to the Netherlands prior to the estimated infection date of the recipient. Pairs were also excluded if the potential transmitter likely was deemed noninfectious,[33] based on a viral load below 200 copies/ml on the estimated infection date of the recipient as derived from LOESS smoothers fitted through longitudinal viral load measurements. Pairs with an estimated time elapsed greater than 16 years were also excluded, since given typical disease progression it is unlikely for two individuals to remain undiagnosed (thus unsampled) for longer than 8 years.[34]

## 2.6 Phylogenetic source attribution

We estimated population-level transmission flows and sources of transmissions from the phylogenetically possible transmission pairs using a Bayesian mixture modelling approach that harnesses both phylogenetic data and time elapsed from the estimated infection date and the sequence sampling date of both the recipient and source, as described previously[20] and in Supplementary Material Section S3.1. Our primary motivation for this approach was to leverage information from estimated infection times, to ascertain that observed patristic distances are consistent with the number of mutations expected under the HIV evolutionary clock. In this approach to source attribution, potential transmission pairs with large patristic distances are not necessarily penalised if the time elapsed between the transmission event and the sampling dates is also large and the observed patristic distances are within the range expected under the HIV evolutionary clock.

The model classifies possible transmission pairs probabilistically to each of two categories represented by components in the model. An observed pair is classified as either being compatible with the signal density corresponding to the HIV evolutionary clock for a true transmission pair, or not, and thus is unlikely an epidemiologically linked pair. We additionally leverage data on age of individuals in each pair to model the mixture weights, representing the probability that a pair belongs to each component. Numerical inference was done with Stan (`cmdstanr` v2.28.1), with 4 chains of 2500 iterations. The model converged with no divergent transitions and fitted the available data well (Supplementary Section S3.2 and Supplementary Figures S3-S4). From the fitted model, we adjusted posterior transmission pair probabilities such that each incident case can have at most one likely source and accounting for the possibility that the true source may have been unsampled. We additionally adjusted for undiagnosed and unsequenced incident cases in 2010-2021 by multiplying transmission pair probabilities with inverse probability sequence sampling weights (Supplementary Material Section S3.3). Population-level transmission flows were then estimated by aggregating over the posterior probabilities that each phylogenetically possible transmission pair was classified as a true transmission pair.

# 3 Results

## 3.1 Increasing frequency of non-B subtypes among Amsterdam MSM

There were 1,335 MSM in the ATHENA cohort who were ever resident in Amsterdam and who had an estimated infection date between $1^{st}$ January 2010 to $31^{st}$ December 2021. Of these, 49% were born in the Netherlands, and 51% were born in other world regions (Table 1 and Figure 1A). Foreign-born Amsterdam MSM had longer median times-to-diagnosis than Dutch-born Amsterdam MSM (10 months [0-6 years] versus 6 months [0-5 years] respectively).[9]

Of the 1,335 Amsterdam MSM with an estimated infection date in 2010-2021, 900 (67%) had a partial HIV *polymerase* sequence available and 824 (92%) were of subtype/CRF B, C, 01AE, A1, 02AG, D, G, and 06cpx (Table 1). The remaining 76 Amsterdam MSM had uncommon subtypes/CRFs or could not be classified. We found sequence sampling fractions were similar across MSM born in different world regions (Supplementary Figure S2).

HIV subtype is a simple indicator into the origin of incident infections; as expected most diagnosed and sequenced Dutch-born Amsterdam MSM with estimated infection date in 2010-2021 had a subtype B virus (371 of 462, 80%) (Figure 1B). More unexpectedly, the large majority of foreign-born Amsterdam MSM also had a subtype B virus (333 of 438, 76%) in 2010-2021, suggesting frequent transmission from Western European or Northern American MSM in who subtype B is most prevalent to both Dutch-born and foreign-born Amsterdam MSM.

To characterize time trends, we next estimated the proportion of non-B incident cases among Dutch-born and foreign-born Amsterdam MSM in each calendar year using sequence data from 2010-2021. Figure 1C shows that the estimated proportion of incident cases with a non-B virus increased over time, even when we assumed that all unsequenced Amsterdam MSM had subtype B. These time trends could indicate changes in transmission dynamics with greater proportions of transmissions from individuals who acquired HIV in locations where subtype B is less prevalent, and/or greater proportions of importations of in-migrating individuals who acquired HIV in locations where subtype B is less prevalent.

## 3.2 Phylogenetic data exclude the majority of potential sources

To identify distinct, ongoing local transmission chains since 2010, we based phylogenetic analyses on a large sequence data set that, in addition to the 900 sequenced Amsterdam MSM with estimated infection in 2010-2021, included a further 2,471 sequences from Amsterdam MSM with estimated infection prior to 2010, 7,028 sequences from individuals enrolled into the ATHENA cohort from the rest of the Netherlands, and 15,806 international sequences from the Los Alamos database closest to the ATHENA sequences (Figure 2). Across the eight subtypes/CRFs, 376 of the 900 sequenced Amsterdam MSM with estimated infection in 2010-2021 were phylogeographically rooted in the background of non-Amsterdam sequences and formed local phylogenetic transmission chains of size 1. These Amsterdam MSM could correspond to importations of new transmission lineages into Amsterdam, or be part of locally ongoing transmission chains in which only one member was observed. In the absence of information on their source cases, we focused on Amsterdam MSM with an estimated infection date in 2010-2021 identified as part of local phylogenetic transmission chains of size $> 1$. For the remaining 524 incident cases, we identified 3,033 Amsterdam MSM who were enrolled into ATHENA with an estimated median infection date prior to that of the incident case. Of those, we excluded the vast majority (99.8% of 1,372,332 potential pairs) based on phylogenetic data; data on death, migration, or viral suppression of the source; and time elapsed exceeding 16 years (Figure 3). Following exclusions, 115 incident cases had no plausible source remaining, leaving for phylogenetic source attribution analysis 2,824 phylogenetically and epidemiologically possible transmission pairs between 409 incident Amsterdam MSM in local phylogenetic transmission chains of size $> 1$ and 742 unique possible sources.

## 3.3 Dutch-born Amsterdam MSM were primary sources of locally acquired infections in 2010-2021

For all Amsterdam MSM who acquired HIV in 2010-2021, we estimated that 56% [CrI 53-58%] of transmissions originated from Dutch-born MSM (Figure 4A). To contextualise these transmission flows, we considered the estimated contribution of Dutch-born MSM to HIV prevalence in Amsterdam MSM over the same time period (Figure 4B). These estimates suggest that Amsterdam MSM born in Suriname & the Dutch Caribbean and Amsterdam MSM born in Eastern & Central Europe were the only population groups that contributed disproportionately to onward transmission over the entire study period, relative to MSM with HIV in Amsterdam from these regions (Figure 4C).

We next characterized the sources of infections to each of the Amsterdam MSM sub-groups stratified by region of birth. In 2010-2021, we estimated that Dutch-born Amsterdam MSM contributed the majority of transmissions to Amsterdam MSM born in the Netherlands, Western Europe, North America & Oceania, South America & the non-Dutch Caribbean, Eastern & Central Europe, and the Middle East & North Africa (Figure 5A, Table 2). The only exception were Amsterdam MSM born in Suriname & the Dutch Caribbean, for whom we estimated that similar proportions of transmission sources were Dutch-born MSM and MSM born in Suriname & the Dutch

Caribbean. We identified a large Amsterdam phylogenetic subgraph with MSM predominantly born in Suriname & the Dutch Caribbean, however additional analyses suggest that this subgraph alone cannot explain our findings and that the higher proportion of within-group transmission sources among Amsterdam MSM born Suriname & the Dutch Caribbean appears more broadly sustained (Supplementary Material S4). Overall, the estimated assortativity coefficient in transmission flows by region of birth was 0.30 [0.28-0.32].

We also estimated the transmission flows across all Amsterdam MSM sub-groups, such that the flows between each sub-group sum to 100% (Figure 5B). In 2010-2021, we estimated that 31% [30-33%] of all transmissions between Amsterdam MSM were between Dutch-born individuals. The second and third largest flows were from Dutch-born MSM to MSM born in other countries in Western Europe, North America & Oceania, and vice versa (9% [8-10%] and 8% [7%-9%], respectively, Table 2).

### 3.4 Foreign-born Amsterdam MSM contribute increasingly to transmission

To further investigate the transmission dynamics underpinning the increasing proportion of non-B subtypes in Amsterdam MSM (Figure 1B-C), we considered each two-year period separately. Sample sizes of incident cases and corresponding phylogenetically and epidemiologically possible sources were limited, with respectively 133 and 1465 in 2010-2011, 91 and 558 in 2012-2013, 81 and 426 in 2014-2015, 57 and 207 in 2016-2017, 36 and 131 in 2018-2019, and 11 and 37 in 2020-2021. Based on the available data, we found that since 2018, foreign-born Amsterdam MSM have been contributing more to local HIV transmission in Amsterdam than Dutch-born MSM in the context of overall declining incidence (Figure 5C).

## 4 Discussion

Amsterdam has been seeing continued and sustained declines in the number of new HIV diagnoses and estimated new infections since 2010, which coincided with the introduction of now well-established test-and-treat strategies, and additional combination interventions introduced by H-TEAM that include pre-exposure prophylaxis (PrEP), innovative test-and-treat strategies at general practioners', the Center for Sexual Health (CSH) of the Public Health Service Amsterdam and hospitals, and studies on motives and barriers for testing.[5] Of the remaining ongoing transmissions, the majority originate from within the city,[9] and around a third are estimated to have been diagnosed within six months of infection,[13] consistent with serological estimates of recency among MSM in Western Europe.[35,36] Here, we integrated pathogen genomic data from Amsterdam, the rest of the Netherlands and from international HIV sequence databases on all primary HIV subtypes and CRFs with demographic and clinical data to estimate the sources of infections in local, ongoing transmission chains among MSM who were ever resident in Amsterdam in 2010-2021. We considered in particular the transmission dynamics of foreign-born Amsterdam MSM, given the slower incidence declines in this group. We found that Dutch-born Amsterdam MSM were the predominant sources of transmission in all Amsterdam MSM populations who acquired their infection locally in 2010-2021, with the exception of Amsterdam MSM born in Suriname & the Dutch Caribbean among whom we estimated that similar proportions of transmissions originated from MSM born in the Netherlands and born in Suriname & the Dutch Caribbean. Overall, these data from Amsterdam provide additional information into the transmission dynamics and infection sources to foreign-born MSM who acquire infection post-migration in Western Europe,[8, 10, 37] and suggest that these foreign-born MSM likely acquired their infection from MSM of the native resident population.

Over time, we found evidence for shifting transmission dynamics towards greater proportions of transmissions originating from foreign-born MSM who are now Amsterdam residents. These shifts in transmission dynamics may be explained by differential uptake of HIV care and prevention services among Amsterdam MSM; for example foreign-born MSM in the Netherlands are less likely to have heard of PEP/PrEP and report experiencing more difficulties accessing healthcare than their native counterparts,[38] and are more likely to present late with qualitative data indicating one reason for this being fear of social stigma associated with a diagnosis.[39]

Our findings should be considered in the context of the following limitations. First, data on the ethnicity of Amsterdam MSM were not available for analysis. It is possible that some of the Dutch-born transmission sources are of non-Dutch ethnicity, and in this case local transmission dynamics among Amsterdam MSM may be more assortative than our analysis suggests. Second, pathogen sequence data were not available for all Amsterdam patients diagnosed with HIV, and we could not estimate the sources of locally acquired infections among Amsterdam residents who acquired HIV through heterosexual contact due to small sample sizes. Considerably more assortative sexual mixing between foreign-born heterosexual individuals has been reported in the Netherlands[40–42] than among MSM, suggesting that transmission dynamics could be markedly different among heterosexuals in Amsterdam as compared to the local transmission dynamics among Amsterdam MSM that we characterise here. Third, while we attempted to account for potential biases due to sequence sampling heterogeneity, these adjustments are based on modeling assumptions and we cannot rule out that missing data may bias our findings. In particular, we cannot exclude the possibility that within-group transmission networks remained unsampled and that the proportion of within-group transmissions among foreign-born Amsterdam MSM is more similar to that seen among Amsterdam MSM born in Suriname & the Dutch Caribbean. Fourth, the possible transmission pairs that we based our inferences on were dependent on estimated infection dates and available demographic and clinical data, and thus their selection carries some uncertainty. We thus conducted several sensitivity analyses (Supplementary Material Section S5), which indicate our primary findings are robust to alterations of our transmission pair selection criteria. Fifth, the sample sizes that underpin our temporal analysis of changes in transmission sources were small, in particular for 2020-2021. Continued genomic surveillance at higher sequence sampling fractions will be essential to substantiate the growing proportions of infections from foreign-born MSM that we identified from the available data.

The city of Amsterdam has already met 2025 UNAIDS 95-95-95 Fast-Track targets,[13] and the focus of all stakeholders at the city level is now converging on reaching zero new HIV infections by 2026.[43] In this effort, detailed phylogenetic analyses can provide helpful information to further target and refine prevention services, and to adapt these in the context of changes in transmission dynamics.[9, 44–46] In Amsterdam, foreign-born MSM continue to be associated with longer times from infection to diagnosis,[9] and are more frequently presenting with a late-stage HIV infection at diagnosis.[47] However, recent programs across the Netherlands have sought to promote early diagnosis, for example through home-based self-testing[48] and developing indicator-condition HIV testing,[49,50] with promising results. More widespread positive and inclusive prevention messaging encouraging knowing one's own HIV status and that of all sexual partners could further raise awareness and prevent local onward transmission among MSM. Our viral phylogenetic analyses indicate considerable potential for these and further strengthened prevention interventions among Amsterdam MSM because the majority of new MSM diagnoses continue to originate from local transmission chains amongst MSM, and growing proportions of infections from foreign-born MSM.

# 5 Funding statement

This study received funding as part of the H-TEAM initiative from Aidsfonds (project number P29701). The H-TEAM initiative is being supported by Aidsfonds (grant number: 2013169, P29701, P60803), Stichting Amsterdam Dinner Foundation, Bristol-Myers Squibb International Corp. (study number: AI424-541), Gilead Sciences Europe Ltd (grant number: PA-HIV-PREP-16-0024), Gilead Sciences (protocol numbers: CO-NL-276-4222, CO-US-276-1712, CO-NL-985-6195), and M.A.C AIDS Fund. The ATHENA database is maintained by Stichting HIV Monitoring and supported by a grant from the Dutch Ministry of Health, Welfare and Sport through the Centre for Infectious Disease Control of the National Institute for Public Health and the Environment. We also gratefully acknowledge funding from the Engineering and Physical Sciences Research Council (PA5083) and the Imperial College Department of Mathematics Cecilia Tanner Research Funding Scheme.

# 6 Author contributions

OR designed the study. OR, PR, GdB oversaw the study. AvS, DB, TvdL oversaw and performed data collection. AB and OR contributed to the analysis. AB and OR wrote the first draft. NP, EGK, LS, AvS, DB, TvdL, MvdV, PR, GdB discussed results and contributed to the revision of the final manuscript.

# 7 Conflict of interest statement

AvS received funding for managing the ATHENA cohort, supported by a grant from the Dutch Ministry of Health, Welfare and Sport through the Centre for Infectious Disease Control of the National Institute for Public Health and the Environment; and received grants unrelated to this study from European Centre for Disease Prevention and Control paid to his institution. PR received grants unrelated to this study from Gilead Sciences, ViiV Healthcare and Merck, paid to his institution; and received Honoraria for lecture from Merck paid to his institution; and received Honoraria from Gilead Sciences, ViiV Healthcare and Merck, paid to his institution. GdB received honoraria to her institution for scientific advisory board participations for Gilead Sciences and speaker fees from Gilead Sciences (2019), Takeda (2018-2022) and ExeVir (2020-current). NP received grants unrelated to this study from ECDC and Gilead Sciences Hellas, paid to his institution; and received honoraria for presentations unrelated to this study from Gilead Sciences Hellas. OR received grants unrelated to this study from the EPSRC, the NIH, and the Bill & Melinda Gates Foundation, paid to his institution.

# 8 Availability of Data and Materials

All code to reproduce the analysis in this paper is available from `github.com/MLGlobalHealth/transmission.flows.by.birthplace`.

|  | | Amsterdam MSM | Sequenced Amsterdam MSM | Amsterdam MSM with estimated infection date in 2010-2021 | Sequenced Amsterdam MSM with estimated infection date in 2010-2021 |
|---|---|---|---|---|---|
|  |  | N (%) | N (%) | N (%) | N (%) |
| Total |  | 6,139 | 3,371 | 1,335 | 900 |
| Age at estimated infection date | 15-24 | 811 (13%) | 466 (14%) | 206 (15%) | 129 (14%) |
|  | 25-34 | 2,108 (34%) | 1,219 (36%) | 483 (36%) | 322 (36%) |
|  | 35-44 | 1,531 (25%) | 919 (27%) | 337 (25%) | 228 (25%) |
|  | 45-54 | 661 (11%) | 413 (12%) | 222 (17%) | 161 (18%) |
|  | 55+ | 212 (3%) | 131 (4%) | 87 (7%) | 60 (7%) |
|  | Unknown | 816 (13%) | 223 (7%) | 0 (0%) | 0 (0%) |
| Region of birth | Netherlands | 3,282 (53%) | 1,933 (57%) | 652 (49%) | 462 (51%) |
|  | W.Europe, N.America & Oceania | 1,019 (17%) | 469 (14%) | 187 (14%) | 109 (12%) |
|  | S. America & non-Dutch Caribbean | 610 (10%) | 292 (9%) | 163 (12%) | 112 (12%) |
|  | Suriname & Dutch Caribbean | 369 (6%) | 247 (7%) | 87 (6%) | 67 (7%) |
|  | Middle East and North Africa | 206 (3%) | 110 (3%) | 79 (6%) | 48 (5%) |
|  | E. & C. Europe | 157 (3%) | 78 (2%) | 65 (5%) | 37 (4%) |
|  | Other | 496 (8%) | 242 (7%) | 109 (8%) | 65 (7%) |
| Subtype | B | - | 3,026 (90%) | - | 704 (78%) |
|  | 01AE | - | 83 (2%) | - | 43 (5%) |
|  | 02AG | - | 68 (2%) | - | 24 (3%) |
|  | A1 | - | 45 (1%) | - | 29 (3%) |
|  | C | - | 38 (1%) | - | 13 (1%) |
|  | G | - | 10 (<1%) | - | 6 (1%) |
|  | D | - | 7 (<1%) | - | 4 (<1%) |
|  | 06cpx | - | 3 (<1%) | - | 1 (<1%) |
|  | Other | - | 97 (3%) | - | 76 (8%) |

Table 1: **Characteristics of Amsterdam MSM with HIV enrolled into ATHENA.**

|  |  | Birth place of incident case | | | | | | |
|---|---|---|---|---|---|---|---|---|
|  |  | Netherlands | W.Europe, N.America & Oceania | Suriname & Dutch Caribbean | S. America & non-Dutch Caribbean | E. & C. Europe | Middle East & North Africa | Other |
| Estimated incident cases | | 673 | 192 | 178 | 91 | 81 | 70 | 114 |
| Number of observed incident cases with sequence data | | 462 | 109 | 112 | 67 | 48 | 37 | 65 |
| Number of phylogenetically and epidemiologically possible sources for incident cases born in each region | | 1391 | 392 | 283 | 293 | 130 | 108 | 227 |
| Sources of Amsterdam MSM transmission for each recipient group, by birth place of source† | Netherlands | 56.8% [53.8-59.8%] | 61.8% [55.5-68.8%] | 38.7% [32.4-45.3%] | 53.7% [44.9-62.3%] | 68.% [58.5-78.5%] | 57.5% [48.8-65.4%] | 39.8% [33.3-50.2%] |
|  | W.Europe, N.America & Oceania | 15% [13.1-16.9%] | 18.1% [12.1-23.8%] | 4.2% [1.3-8.1%] | 9.1% [3.7-15.7%] | 11.2% [4.5-20.3%] | 3.1% [0.5-6.7%] | 32.9% [15.3-40.6%] |
|  | Suriname & Dutch Caribbean | 9.1% [7.3-11.1%] | 6.2% [3-9.8%] | 36.1% [29.3-42.9%] | 6.4% [1.8-11.7%] | 4.1% [0.5-10%] | 3.2% [0.5-7.8%] | 5.6% [1.6-12.4%] |
|  | S. America & non-Dutch Caribbean | 5.8% [4.3-7.4%] | 5.9% [2.1-9.2%] | 7.5% [4.2-12.4%] | 12.6% [5.9-19.4%] | 5.4% [0.7-13.3%] | 5.6% [2.2-9.4%] | 6.9% [2.2-14.1%] |
|  | E. & C. Europe | 4.3% [3.3-5.5%] | 5.2% [2.6-8.2%] | 0.1% [0.0-2.3%] | 7.3% [4.4-10.6%] | 0.0% [0.0-2.0%] | 12.8% [5.2-18.5%] | 0.9% [0.0-6.9%] |
|  | MENA | 3.5% [2.4-4.7%] | 1.3% [0.2-3.2%] | 3.6% [1.5-6.8%] | 0.8% [0-4.2%] | 0% [0-1.4%] | 12.2% [7.3-18.0%] | 2.0% [0.4-5.5%] |
|  | Other | 5.4% [4.0-7.1%] | 1.1% [0.1-3.3%] | 9.1% [5.6-12.2%] | 9.5% [3.5-15.3%] | 8.9% [3.7-15.4%] | 5.3% [2.3-8.8%] | 11.1% [6.8-15.5%] |
| Amsterdam MSM transmission flows, by birth place of source‡ | Netherlands | 31.3% [29.6-33.1%] | 8.7% [7.7-9.7%] | 3.0% [2.4-3.5%] | 3.8% [3.0-4.6%] | 3.7% [3.0-4.4%] | 2.5% [2.0-3.0%] | 2.5% [2.1-3.0%] |
|  | W.Europe, N.America & Oceania | 8.2% [7.2-9.3%] | 2.5% [1.6-3.4%] | 0.3% [0.1-0.6%] | 0.6% [0.3-1.1%] | 0.6% [0.2-1.1%] | 0.1% [0.0-0.3%] | 2.1% [0.9-2.7%] |
|  | Suriname & Dutch Caribbean | 5% [4.0-6.1%] | 0.9% [0.4-1.4%] | 2.8% [2.1-3.4%] | 0.5% [0.1-0.9%] | 0.2% [0.0-0.5%] | 0.1% [0.0-0.3%] | 0.3% [0.1-0.7%] |
|  | S. America & non-Dutch Caribbean | 3.2% [2.4-4.1%] | 0.8% [0.3-1.3%] | 0.6% [0.3-0.9%] | 0.9% [0.4-1.4%] | 0.3% [0.0-0.8%] | 0.2% [0.1-0.4%] | 0.4% [0.1-0.8%] |
|  | E. & C. Europe | 2.4% [1.8-3.0%] | 0.7% [0.4-1.1%] | 0.0% [0.0-0.2%] | 0.5% [0.3-0.7%] | 0.0% [0.0-0.1%] | 0.6% [0.2-0.8%] | 0.1% [0.0-0.4%] |
|  | MENA | 1.9% [1.3-2.6%] | 0.2% [0.0-0.5%] | 0.3% [0.1-0.5%] | 0.1% [0.0-0.3%] | 0.0% [0.0-0.1%] | 0.5% [0.3-0.8%] | 0.1% [0.0-0.3%] |
|  | Other | 3.0% [2.2-3.9%] | 0.1% [0.0-0.5%] | 0.7% [0.4-0.9%] | 0.7% [0.2-1.1%] | 0.5% [0.2-0.8%] | 0.2% [0.1-0.4%] | 0.7% [0.4-0.9%] |

† columns sum to 100%
‡ rows and columns sum to 100%

Table 2: **Estimated total incident cases among Amsterdam MSM in 2010-2021 and sources locally acquired infections in Amsterdam transmission chains, stratified by place of birth.**

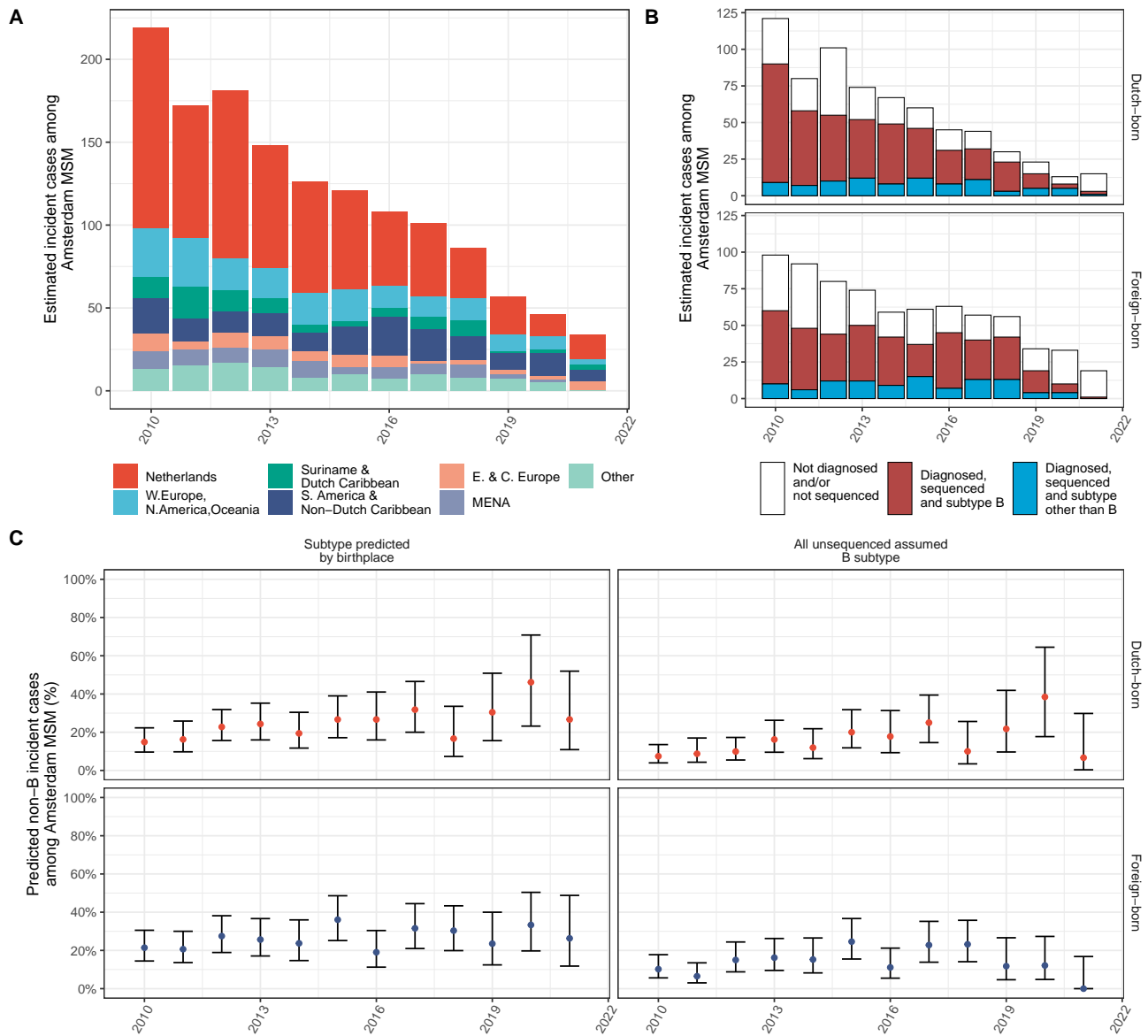

Figure 1: **Time trends in HIV subtype frequencies among Amsterdam MSM**. A) Estimated number of HIV incident Amsterdam MSM by calendar year of their median infection time estimates, stratified by place of birth. B) Estimated number of HIV incident Amsterdam MSM by calendar year of their median infection time estimates, and stratified by HIV subtype where available. C) Predicted proportion of non-B subtypes among HIV incident Amsterdam MSM by calendar year of median infection time estimates, accounting for unsequenced and undiagnosed individuals under two possible scenarios. The left scenario assumes that the acquisition of subtypes among unsequenced cases reflects proportions in sequenced cases by place of birth. The right scenario assumes that all unsequenced cases acquired a subtype B virus. Throughout, central estimates (dots) are shown along 95% Agresti-Coull confidence intervals (error bars).

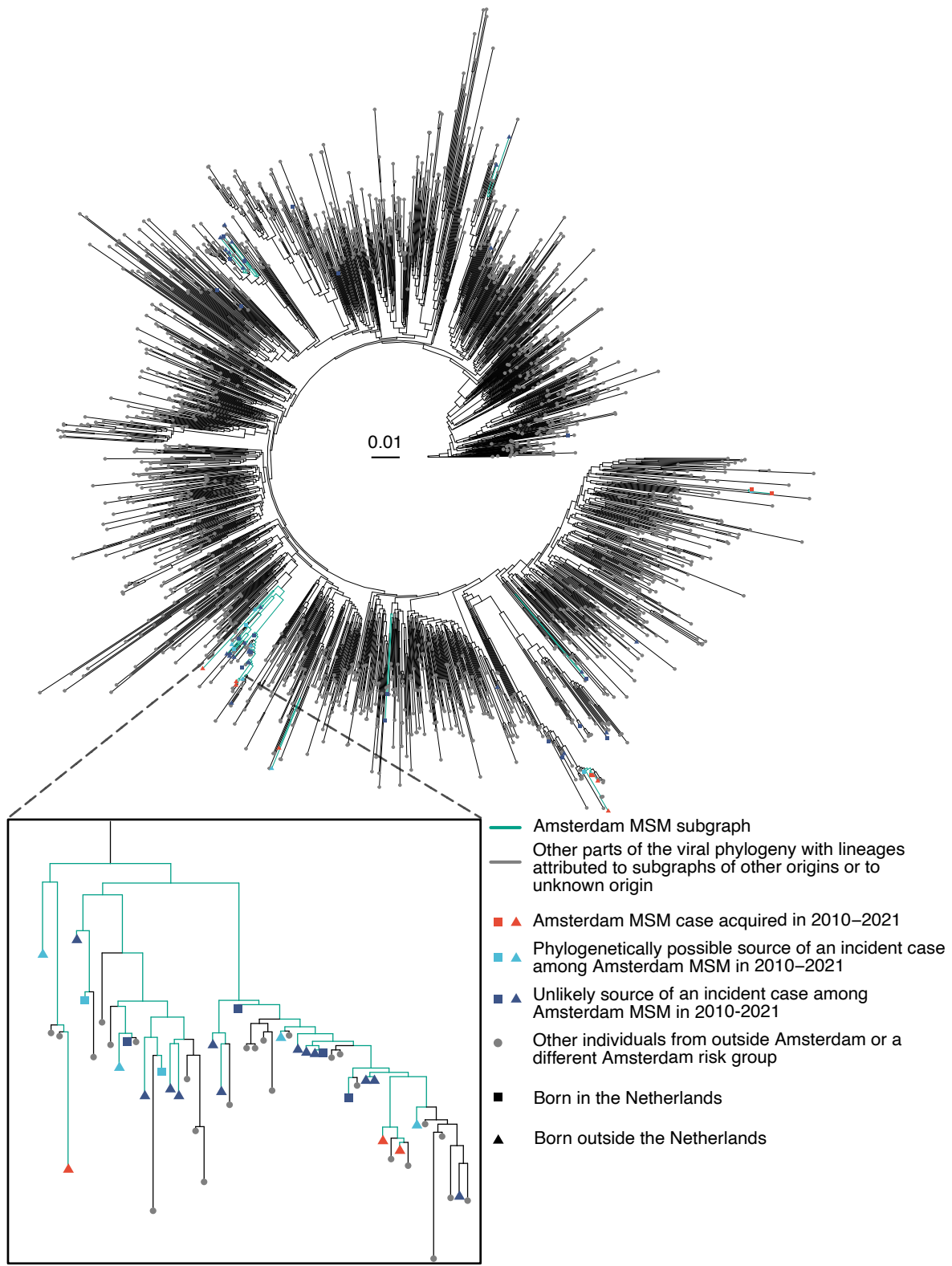

Figure 2: **Phylogenetic tree for Amsterdam MSM with circulating recombinant form 02AG and international background sequences.** Amsterdam MSM subgraphs in the phylogenetic tree were reconstructed with phyloscanner and are coloured by continuous green branches. Red tips denote incident cases among Amsterdam MSM estimated to have acquired their infection in 2010-2021. Light blue tips denote phylogenetically possible sources of each incident case, dark blue tips denote Amsterdam MSM who are unlikely sources to these incident cases, given phylogenetic, epidemiological and clinical data. Grey tips denote non-Amsterdam or non-MSM cases. Squares represent Amsterdam MSM born in the Netherlands and triangles are those with a migration background.

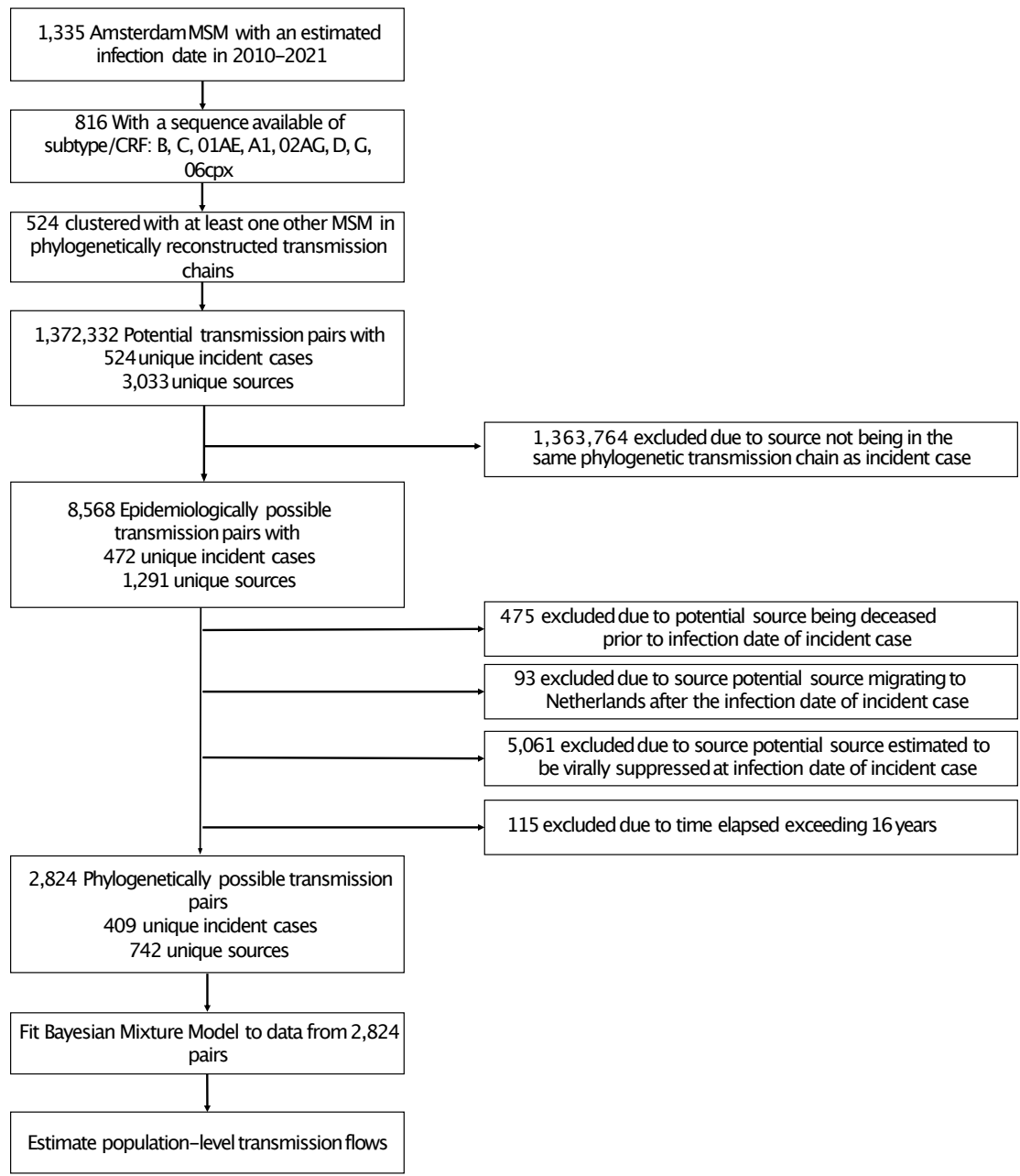

Figure 3: **Flowchart for the selection of phylogenetically and epidemiologically possible transmission pairs to Amsterdam MSM who we estimate acquired infections in local transmission chains between 2010-2021**.

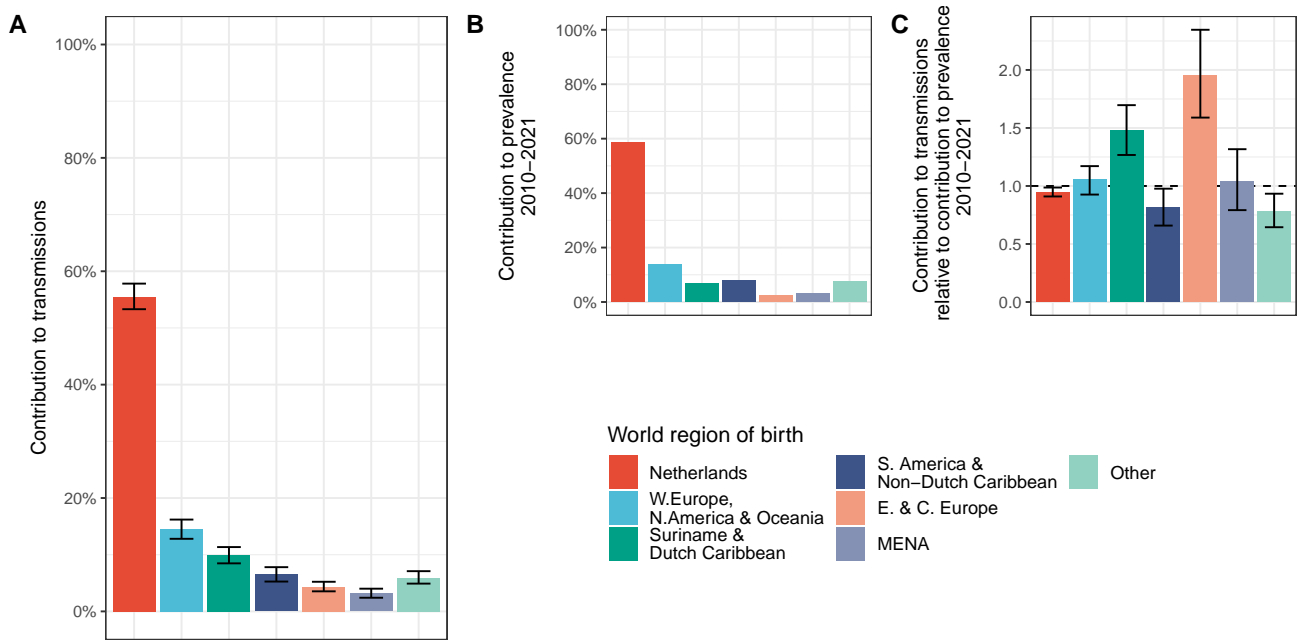

Figure 4: **Sources of infections in Amsterdam MSM who locally acquired infection in Amsterdam transmission chains in 2010-2021.** **A)** Estimated contributions of Dutch-born and foreign-born Amsterdam MSM to transmission to Amsterdam MSM in 2010-2021, with sources stratified by region of birth. Posterior median estimates (bars) are shown along posterior 95% credible intervals (error bars). **B)** Estimated contributions of Dutch-born and foreign-born Amsterdam MSM to HIV prevalence among Amsterdam MSM in 2010-2021. **C)** Estimated contribution of Amsterdam MSM groups to transmission relative to their contribution to people with HIV in 2010-2021.

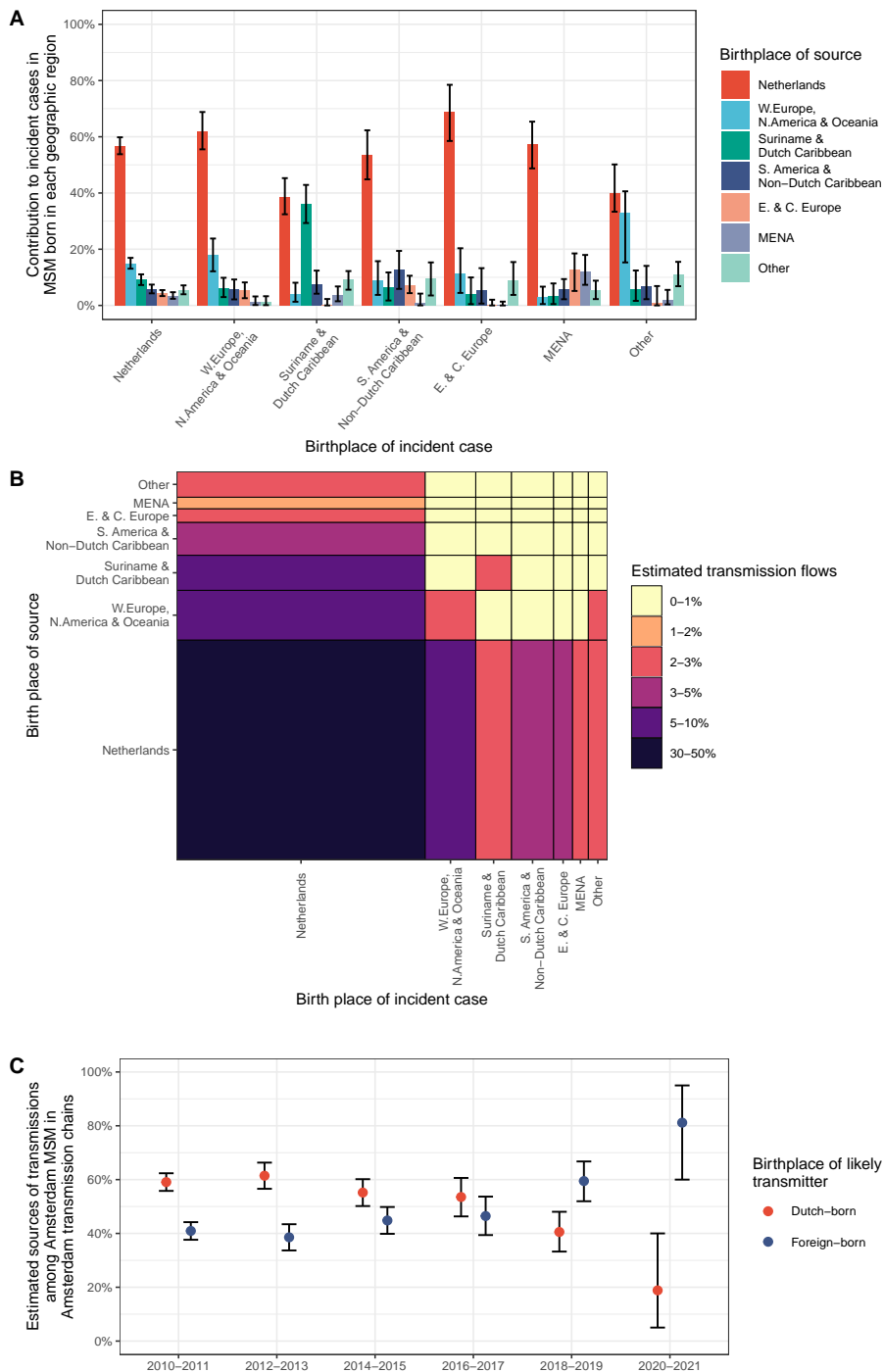

Figure 5: **Sources of infections in Amsterdam MSM who acquired a locally acquired infection in Amsterdam transmission chains in 2010-2021, stratified by region of birth of both sources and incident cases and over time.** **A**) Sources of infection in Amsterdam MSM who acquired infection through Amsterdam MSM transmission chains in 2010-2021. Region of birth of incident cases are shown on the x-axis, and the estimated contribution of transmission sources within each incident group in colour. Posterior median estimates (bars) are shown along with posterior 95% credible intervals (error bars). **B**) Estimated transmission flows between Amsterdam MSM sub-groups in 2010-2021. The estimated posterior median proportion of transmission flows from the source group to the recipient group are visualised in colours. Cell widths correspond to the contribution of each Amsterdam MSM group to incidence, and cell heights correspond to the contribution of each Amsterdam MSM group to transmission sources. **C**) Sources of infection in Amsterdam MSM who acquired infection through Amsterdam MSM transmission chains in 2010-2021 by two-year period. Year of acquired infection are shown on the x-axis in two-year intervals, and the estimated contribution of transmission sources for incident cases among Dutch-born and foreign-born MSM are shown in colour. Posterior median estimates (dots) are shown along with posterior 95% credible intervals (error bars).

Supplementary Text to

# Sources of HIV infections among MSM with a migration background: a viral phylogenetic case study in Amsterdam, the Netherlands

Blenkinsop et. al.

# Contents

# Supplementary Figures

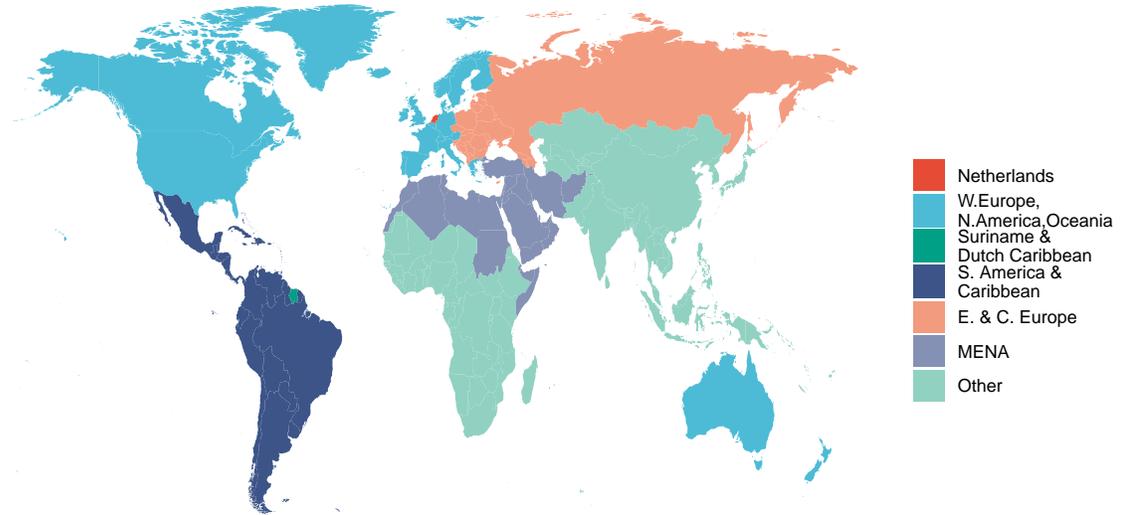

Figure S1: **Geographic regions for primary migrant groups among Amsterdam MSM.**

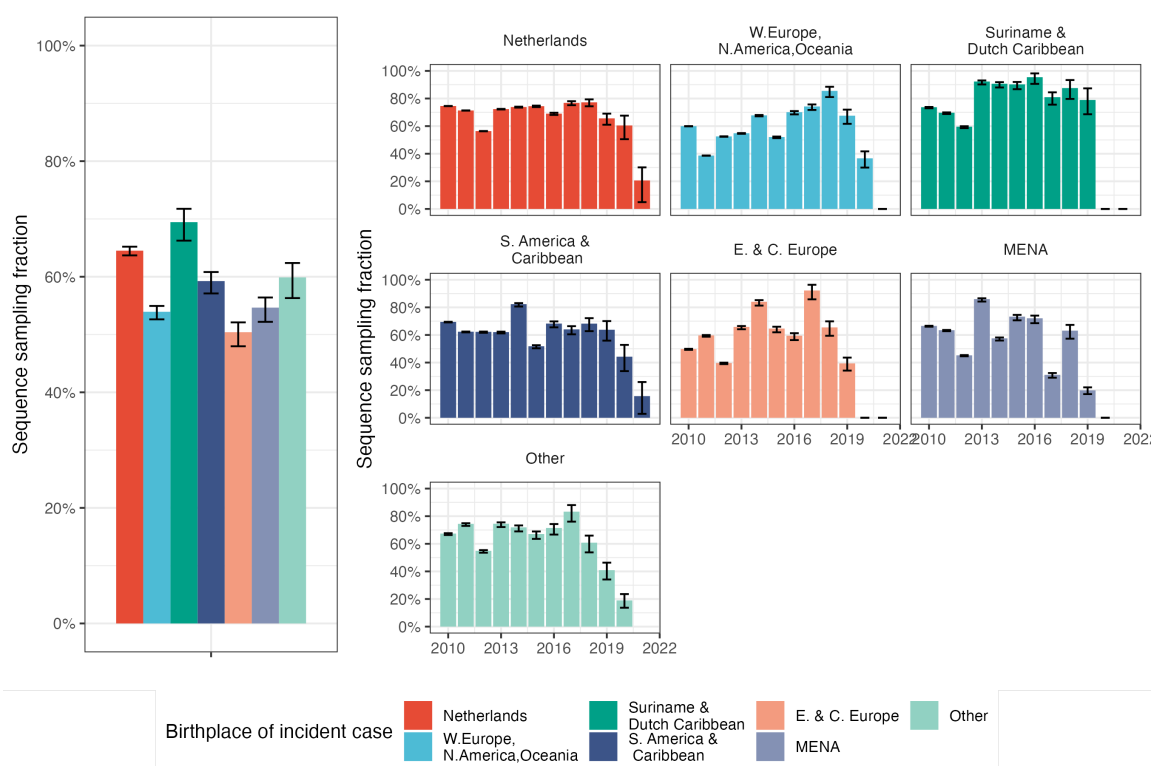

Figure S2: **Sequence sampling fractions by year, stratified by place of birth.** Proportion of estimated incident cases with a sequence available, by year of estimated HIV acquisition, with 95% credible intervals. The number of incident cases were estimated using bivariate linear mixed model (see text).

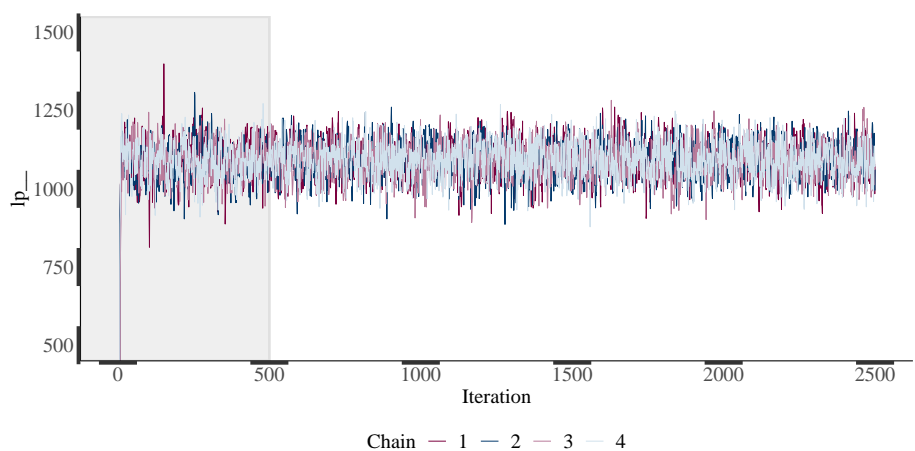

Figure S3: **Trace plot of parameter with the smallest effective sample size for the Bayesian Mixture Model used in the central analysis.**

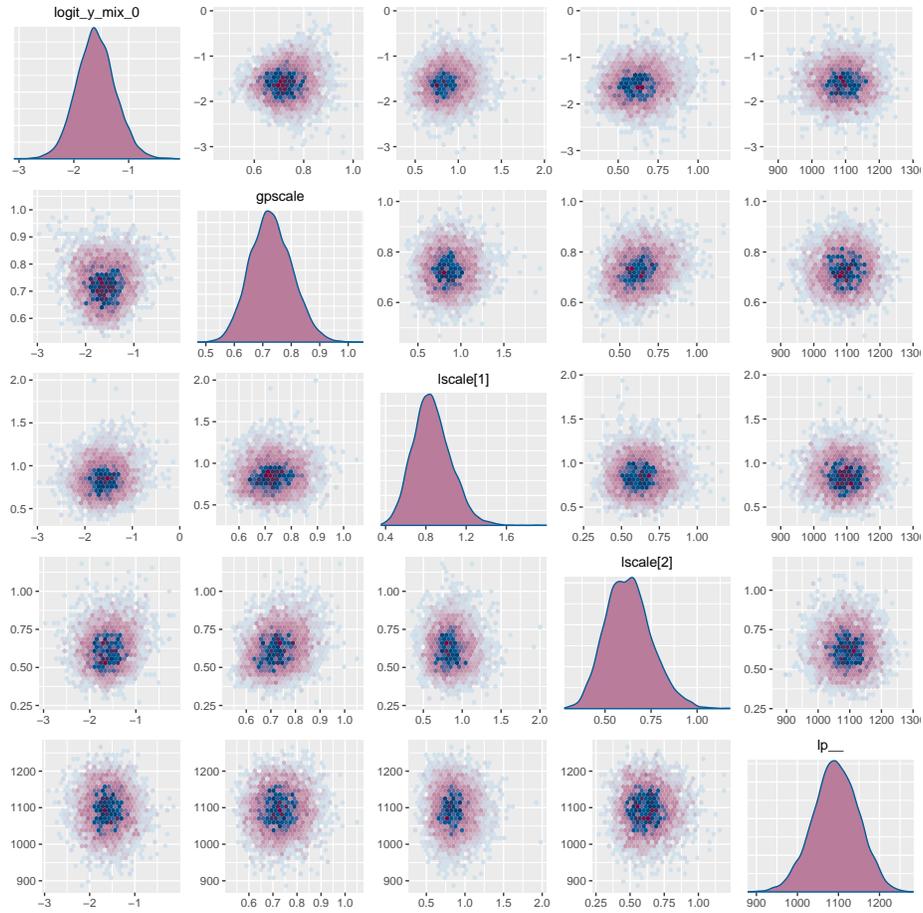

Figure S4: **Pairs plot of the estimated joint posterior density of all model parameters of the Bayesian Mixture Model used in the central analysis.** Coloured hexagons represent the binned 2D counts of posterior draws across monte carlo samples.

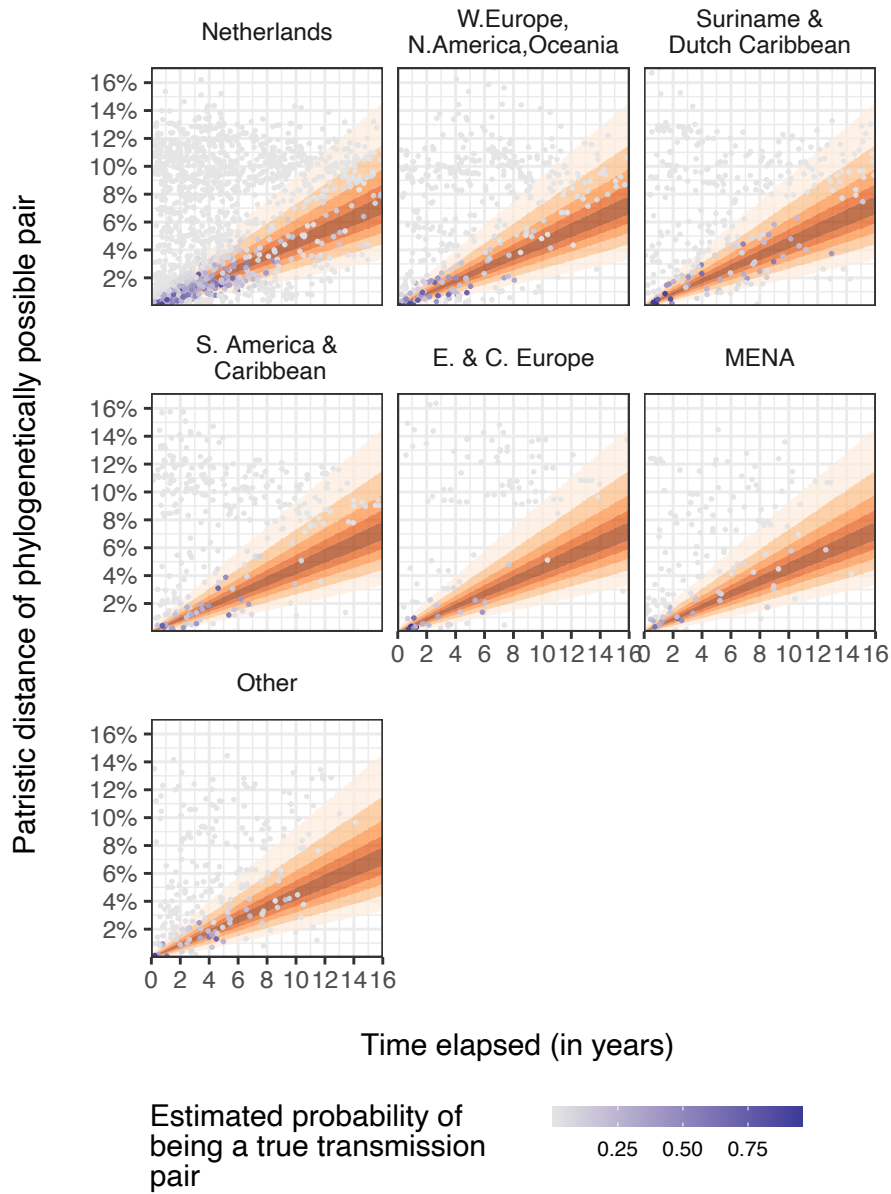

Figure S5: **Posterior median transmission pair probabilities for all phylogenetically possible pairs, by region of birth of the potential source in each pair.** Each point represents the patristic distance and estimated time elapsed of each phylogenetically possible transmission pair, and is coloured according to the posterior median transmission pair probability estimated by the model. Facets represent the geographic region of birth of the phylogenetically possible source in each pair. The orange bands represent the posterior quantiles of the evolutionary clock model, fitted to data from independent data on confirmed transmission pairs in Belgium (see text).

# S1 Estimating HIV prevalence among Amsterdam MSM

To estimate population-level denominators for contextualising transmission flows, we considered all ATHENA participants since the start of data collection who ever resided in an Amsterdam postcode and self-reported as MSM, unless specified otherwise.

In order to obtain estimates of HIV prevalence among Amsterdam MSM stratified by place of birth, we followed methods previously described [1]. Specifically, we fitted a hierarchical Bayesian Weibull likelihood model to time since infection estimates by geographic place of birth for a sub-cohort of Amsterdam MSM estimated to have acquired HIV in 2010-2015, who are least likely subject to right censoring bias. Here, we stratified ATHENA study participants by geographic regions of birth, $\mathcal{K}$ = {Western Europe, North America & Oceania, Eastern and Central Europe, Suriname & the Dutch Caribbean, South America & the non-Dutch Caribbean, Middle East & North Africa (MENA), Other}; see also Supplementary Figure S1.

From the fitted model, we estimated the number of incident cases among Amsterdam MSM born in geographic region $k \in \mathcal{K}$ acquired in year $y \in \mathcal{Y} = \{1996,\ldots,2021\}$, following previous methods [1], denoted by $N_{ky}^{\mathrm{I}}$. We then estimated the number of people with HIV (PWH) in year $y$ for MSM born in region $k$ by summing over historical years since the start of ATHENA,

$$N_{ky}^{\mathrm{PWH}} = \sum_{i=1}^{y} N_{ki}^{\mathrm{I}} - N_{ki}^{\mathrm{died}}, \tag{S1}$$

where $N_{ki}^{\mathrm{died}}$ are the number of individuals reported to have died in year $i$.

We calculated the contribution of MSM born in each geographic region $k$ towards total prevalence over 2010-2021 as a weighted average as follows,

$$\pi_k^{\mathrm{prevalence}} = \sum_{i=2010}^{2021} \frac{N_{ki}^{\mathrm{PWH}}}{\sum_{k \in \mathcal{K}} N_{ki}^{\mathrm{PWH}}} \omega_i, \tag{S2}$$

where $\omega_i = \frac{\sum_k N_{ki}^{\mathrm{PWH}}}{\sum_k \sum_j N_{kj}^{\mathrm{PWH}}}$ are the weights corresponding to the estimated total number of PWH among MSM in Amsterdam for year $i \in \mathcal{I} = \{2010,\ldots,2021\}$.

## S2 Distribution of subtypes among Amsterdam MSM

To assess whether the proportion of non-B subtypes has increased over time, we begin with the observed proportions from those MSM with a sequence (Supplementary Figure S6). To make population-level inferences, we estimated the subtypes of the unsequenced and/or undiagnosed MSM, based on the estimated total incident cases. Generally, the number of sequenced ($N_{kis}^{\text{Seq}}$) and unsequenced ($N_{kis}^{\text{U}}$) individuals sum to the total estimated incident cases,

$$N^{\text{I}} = \sum_{kis}(N_{kis}^{\text{Seq}} + N_{kis}^{\text{U}}), \tag{S3}$$

where $k \in \mathcal{K}$ denotes geographic region of birth, $i \in \mathcal{Y}$ denotes year of acquisition, and $s \in \mathcal{S} =$ {B, non-B} denotes HIV subtype. We used this to estimate the total number of unsequenced individuals among new cases in 1996-2021 by summing over the indices of $N_{kis}^{\text{U}}$.

In the most conservative scenario, we assume all unsequenced and/or undiagnosed MSM acquired a subtype B virus,

$$N_{kis}^{\text{U}} = \begin{cases} N_{ki}^{\text{I}} - \sum_{ki} N_{kis}^{\text{Seq}} & \text{when } s = \text{B} \\ 0 & \text{when } s = \text{Non-B}. \end{cases} \tag{S4}$$

We then asses whether we continue to observe an increase in the proportion of non-B viruses over time based on the total number of incident cases of subtype non-B ($N_{kis}^{I}$).

Alternatively, we may assume that the HIV subtypes of unsequenced and/or undiagnosed MSM follows similar trends to those observed in the sampled MSM. For this, we apply the proportion of MSM of subtype B, stratified by place of birth $k$ and year $i$, denoted by $\pi_{ki \text{ non-B}}$, to the total MSM living with HIV minus those sequenced,

$$N_{kis}^{\text{U}} = \pi_{ki \text{ non-B}}(N_{ki}^{\text{I}} - N_{kis}^{\text{Seq}}). \tag{S5}$$

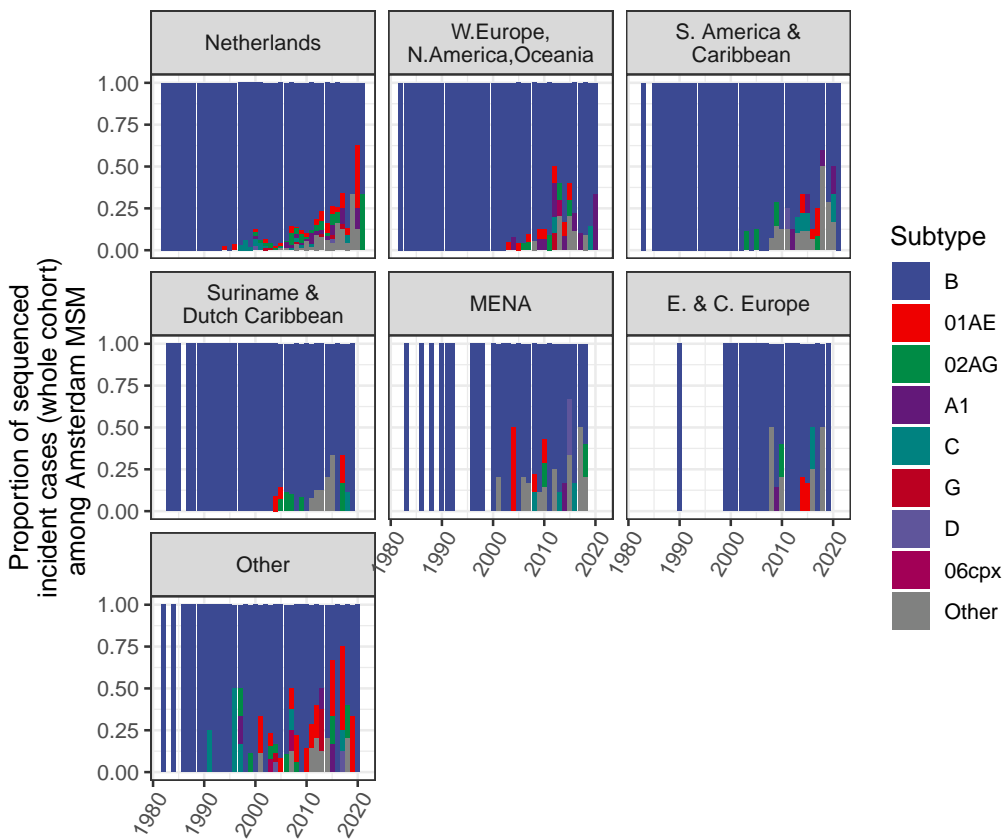

Figure S6: **Proportion of sequenced incident cases among Amsterdam MSM estimated to have been acquired from 1980 to 2021, by geographic region of birth and HIV subtype.**

## S3 Estimating transmission flows

### S3.1 Bayesian Mixture Model

We follow methods described in detail in previous work [2], and here provide only a summary of the approach the Bayesian mixture model approach for phylodynamic source attribution.

**Data** The setup, as outlined in main text, consists of phylogenetically and epidemiologically possible pairs of MSM, who are inferred to belong to the same local transmission network and have epidemiological and clinical data consistent with HIV transmission. Each pair is summarised by a two-dimensional vector comprising the time elapsed and the patristic distance between two individuals in the inferred phylogeny. The former is calculated as the cumulative time elapsed between the estimated infection date of the incident case and the sequence sampling date of both the source and recipient. The latter quantity is calculated as the sum of the branch lengths connecting two individuals in the tree, corresponding to the number of nucleotide substitutions between them.

**Estimating the evolutionary clock of HIV-1** The central idea of the mixture model is to classify observations (in this case potential transmission pairs) into two distinct densities, a signal component which characterises the data for a true transmission pair and a background component, which characterises the data for pairs which are unlinked epidemiologically. The density for the latter is straightforward, since unlinked pairs should exhibit no association between time elapsed and patristic distance. For the density corresponding to the true transmission pairs, it is necessary to know the expected genetic diversity between two individuals, given the time elapsed between them, corresponding to the evolutionary clock for HIV-1. We trained the signal component of the model on data from known transmission pairs to learn this relationship between time and genetic diversity. We leveraged published data from a study in Belgium in which a phylogenetic tree was validated with known transmission history for epidemiologically confirmed transmission pairs [3]. Following methods in [2], we fitted a hierarchical model with a gamma likelihood to the time elapsed and patristic distances from the Belgian study, estimating random effects for each unique pair of individuals.

**Two-component mixture model** We developed a two-component Bayesian hierarchical mixture model, similar to those used widely for classification problems [4]. We embedded the gamma model within the signal component of the mixture model, fixing its hyper-parameters to the medians of their posterior predictive distribution from the evolutionary clock model fitted to the Belgian data and re-estimating the random effects for the new unseen pairs in the model from Amsterdam MSM. We assumed a 2D uniform distribution for the background component over time elapsed and patristic distance, with no parameters to be estimated. The model describes the likelihood that a particular combination of a patristic distance and time elapsed arose by chance through a background density or is compatible with a signal density, the evolutionary clock. In the mixture model, each observation has a probability of belonging to the each of the two components, given by an unknown parameter to be estimated, in which the probabilities for the signal and background component must sum to one. We allowed the mixture probability to vary for each pair, and the probability of belonging to the signal component was modelled with a linear predictor, to estimate the probability a pair represents a true transmission pair given additional covariates from the incident case and putative source in each case. There is typically an association between ages of partners among MSM [5, 6], so the bivariate ages of the source and recipient in a pair is likely informative of being a true transmission pair. We therefore incorporated the age of both individuals within each pair on the estimated date of transmission to the incident case into the linear predictor for the pair-specific mixture parameters. Finally, we fitted the mixture model to the time elapsed, patristic distances and ages corresponding to each plausible pair of Amsterdam MSM.

### S3.2 Bayesian inference

The model was fitted with `cmdstanr` v.2.28.1, with 4 chains of 2500 samples each, including a burn-in of 500. The model converged and mixed well, with a smallest effective sample size across the parameters of 3492, largest Rhat of 1.001 and no divergences.

### S3.3 Target quantities

**Probability that a phylogenetically observed pair represents a truly linked transmission pair.** From the joint posterior, we estimated the probability that each phylogenetically possible pair with source $i$ and recipient $j$ belongs to the signal component of the mixture and thus represents a truly linked pair $\{c = 1\}$ (denoted $p_{ij}^{c=1}$). The probability is specified such that whilst each incident case, $j$, may have multiple phylogenetically possible sources, $i_1, \ldots, i_{L_j}$, no other point involving $j$ can be simultaneously classified as a true transmission pair. The model also has the property that for any $j$ the sum $\sum_i p_{ij}^{c=1}$ can be close to zero and never exceeds one, meaning the model can account for event the true source may not be among any of the observed, phylogenetically possible transmission pairs involving $j$. The probability a pair $i_u, j$ represent a true pair is therefore estimated by,

$$
\begin{aligned}
\rho_{i_u,j} | \boldsymbol{X} = \\
\left( \omega_{i_u,j} \, p(D_{i_u j} | T^e_{i_u j}, Z_{i_u j} = 1) \prod_{v \neq u} (1 - \omega_{i_v,j}) \, p(D_{i_v j} | T^e_{i_v j}, Z_{i_v j} = 0) \right) \Big/ \\
\left[ \sum_{w=1}^{n^{\mathcal{P}}_j} \left( \omega_{i_w,j} \, p(D_{i_w j} | T^e_{i_w j}, Z_{i_w j} = 1) \prod_{v \neq w} (1 - \omega_{i_v,j}) \, p(D_{i_v j} | T^e_{i_v j}, Z_{i_v j} = 0) \right) + \right. \\
\left. \prod_{v=1}^{n^{\mathcal{P}}_j} (1 - \omega_{i_v,j}) \, p(D_{i_v j} | T^e_{i_v j}, Z_{i_v j} = 0) \right].
\end{aligned}
\tag{S6}
$$

**Transmission flows adjusted for incomplete sampling of incident cases.** It is possible that not all incident cases since 2010 are sampled. To adjust for these, we first denote a partition of the study population with $\mathcal{A}$, and population groups in this partition by $a, b \in \mathcal{A}$. The number of incident cases born in geographic region $a$ with an estimated date of HIV acquisition in year $y \in \mathcal{Y} = \{2010, \ldots, 2021\}$ is denoted by $N^D_{ay}$. Using time since infection estimates as previously described, we estimated the proportion of individuals who were undiagnosed by the end of follow-up at the start of 2022, and correspondingly denote the total number of incident cases, including those undiagnosed, by $N^I_{ay}$. The number of individuals who were diagnosed and have a sequence available is given by $N^S_{ay}$. We define the sequence sampling probability of each incidence case by geographic region of birth, $a$, by $\xi_{ay} = \frac{N^S_{ay}}{N^I_{ay}}$. We first estimated the population-level transmission counts originating from MSM born in region $a$ to MSM born in region $b$,

$$
Z_{ab} = \sum_{i \in a, j \in b} Z_{ij} = \sum_{i \in a} \sum_{j \in b} \frac{\rho_{ij}}{\xi_{bt(j)}}
\tag{S7}
$$

for all $a, b \in \mathcal{A}$, where $t(j)$ is the estimated year of HIV acquisition of recipient $j$. For subgroups in which $N^S_{bt} = 0$ but $N^I_{bt} > 0$, we calculate $\xi_{bt}$ as $\frac{0.1}{N^I_{bt}}$, to allow estimates of $Z_{ab}$ to be greater than zero in the event we did not sample any MSM. From this we calculate the population transmission flows originating from MSM born in region $a$ through,

$$
\delta_a = \left( \sum_{b \in \mathcal{A}} Z_{ab} \right) \Big/ \left( \sum_{c,d \in \mathcal{A}} Z_{cd} \right),
\tag{S8}
$$

such that $\sum_a \delta_a = 1$. Just considering MSM born in a single world region in the denominator, the proportion of transmission flows among MSM born in region $b$ originating from MSM born in region $a$ is given by,

$$\delta_{ab} = Z_{ab} \bigg/ \bigg( \sum_{c \in \mathcal{A}} Z_{cb} \bigg), \tag{S9}$$

such that $\sum_a \delta_{ab} = 1$. Finally, the population-level flows from MSM born in region $a$ to MSM born in region $b$ out of all transmissions is given by,

$$\pi_{ab} = Z_{ab} \bigg/ \bigg( \sum_{c,d \in \mathcal{A}} Z_{cd} \bigg), \tag{S10}$$

such that $\sum_a \sum_b \pi_{ab} = 1$.

**Transmission flows relative to prevalence.** To contextualise the transmission flows by their relative population sizes, we calculate first the contribution of MSM born in each world region to prevalence as a weighted sum over the years 2010-2021. The weights for the years are calculated as,

$$\omega_y = \frac{\sum_{a \in \mathcal{A}} N^{\text{PWH}}_{ay}}{\sum_{y=2010}^{2021} \sum_{a \in \mathcal{A}} N^{\text{PWH}}_{ay}}. \tag{S11}$$

Then, the contributions of each world region is,

$$p_a = \sum_{y=2010}^{2021} \omega_y \frac{N^{\text{PWH}}_{ay}}{\sum_{a \in \mathcal{A}} N^{\text{PWH}}_{ay}}. \tag{S12}$$

Then the relative flows are calculated as,

$$\phi_a = \frac{\delta_a}{p_a}, \tag{S13}$$

for each Monte Carlo sample.

**Newman's assortativity coefficient.** We also quantified the degree of mixing in transmissions occurring between MSM born in different geographic regions through Newman's assortativity coef-ficient [7]. This is calculated as,

$$r = \frac{\sum_{a\in\mathcal{A}} \delta_{aa} - \sum_{a\in\mathcal{A}}\sum_{b\in\mathcal{A}} \delta_{ab}^2}{1 - \sum_{a\in\mathcal{A}}\sum_{b\in\mathcal{A}} \delta_{ab}^2}. \tag{S14}$$

## S4 Additional analyses of transmission dynamics among Amsterdam MSM born in Suriname & the Dutch Caribbean

We estimated in the central analysis that among Amsterdam MSM born in Suriname & the Dutch Caribbean as many transmissions originated from other MSM born in the same region as from Dutch-born MSM.

We investigated this further and identified there was one large phylogenetic subgraph with 11 MSM born in Suriname & the Dutch Caribbean, of who five were incident cases acquired since 2010. Excluding this subgraph, we re-estimated the transmission flows towards MSM born in Suriname & the Dutch Caribbean and found that 28% of transmissions were still attributed to have their source in other MSM born in Suriname & the Dutch Caribbean. This corresponded to a reduction of 8% in estimated transmission flows, and suggest that the identified large phylogenetic subgraph alone does not explain the estimated, increased within-group transmission flows among Amsterdam MSM born in Suriname & the Dutch Caribbean.

We then considered all phylogenetically possible within-group transmission pairs, and summed the density values of the mixture model component that captures the likelihood of each observed pair being a true transmission pair as an empirical measure of evidence of within-group transmission flows (Supplementary Figure S7). The analysis indicated that MSM born in Suriname & the Dutch Caribbean had the second most frequent counts after Dutch-born pairs, suggesting there were many phylogenetically and epidemiologically plausible pairs between MSM born in this region that are compatible with direct transmission between them, and confirming that the reconstructed phylogenetic subgraph alone did not explain the estimated, increased within-group transmission flows among Amsterdam MSM born in Suriname & the Dutch Caribbean.

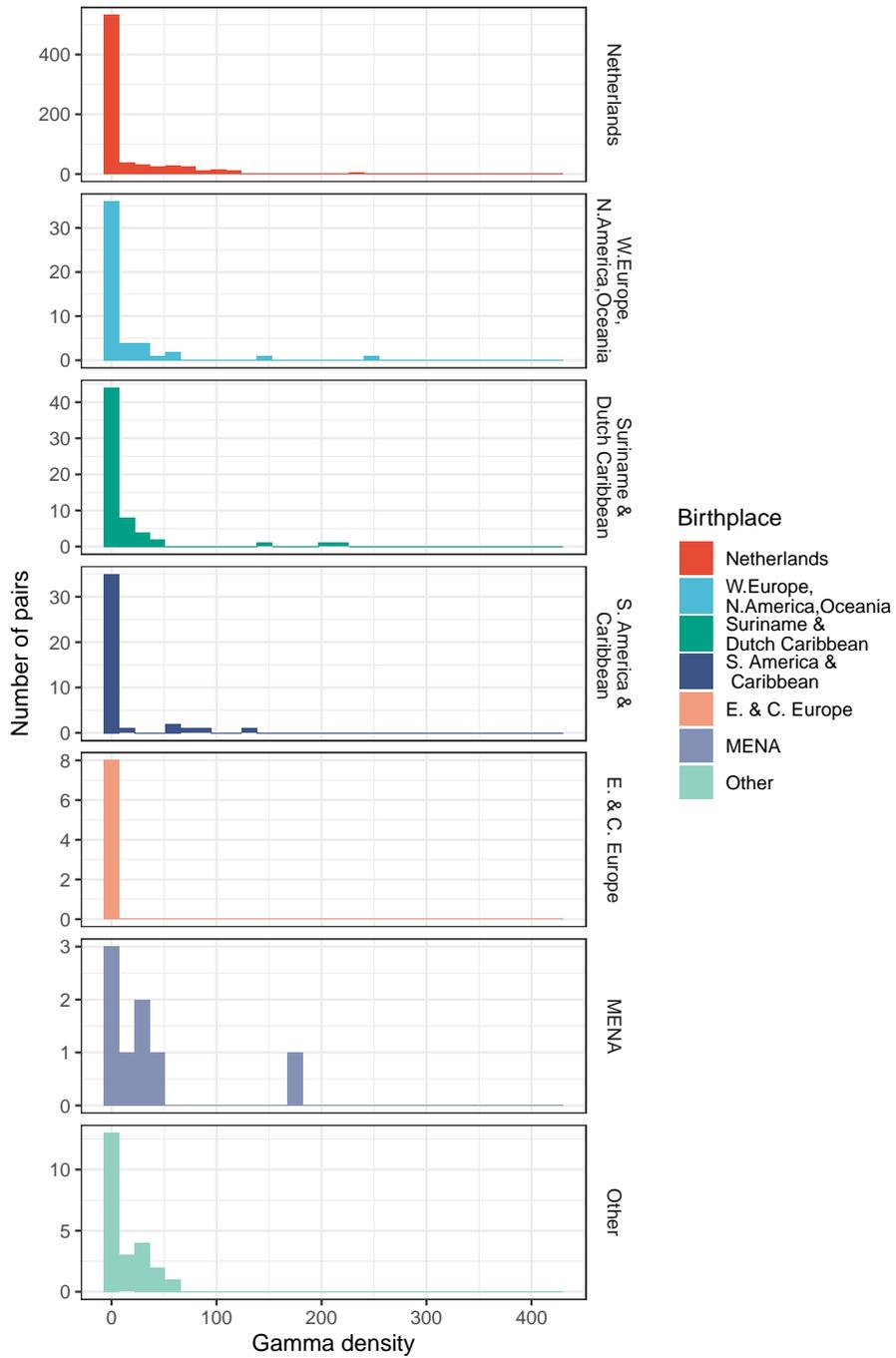

Figure S7: **Histogram of the density values of the mixture model component that captures the likelihood of each observed pair being a true transmission pair gamma densities, for phylogenetically possible transmission pairs with both individuals born in the same geographic region.**

## S5 Sensitivity analyses

We performed a series of sensitivity analyses to investigate the impact of each of the exclusion criteria shown in Figure 3 on the transmission flow estimates. Specifically, we re-fitted the model omitting each exclusion criteria in turn and estimated sources of infection for the revised set of potential transmission pairs. Table S1 compares the estimated transmission flows among Amsterdam MSM originating from each of the seven geographic regions when the model was fitted to pairs applying all exclusion criteria (the central analysis), and by omitting each exclusion criteria using epidemiological and clinical data in turn. Across the sensitivity analyses, we observed similar proportions of transmissions that originated from Amsterdam MSM born in the different geographic regions considered, suggesting in turn that our findings are robust with regards to the additional exclusion criteria based on epidemiologic data that we used in this analysis.

|  |  | Estimated transmission flows originating from MSM born in each region | | | | | | |
| --- | --- | --- | --- | --- | --- | --- | --- | --- |
| Exclusion criteria omitted | Number of pairs | Netherlands | W.Europe, N.America, Oceania | Suriname & Dutch Caribbean | S. America & non-Dutch Caribbean | E. & C. Europe | Middle East & N. Africa | Other |
| None (central analysis) | 2,824 | 56% | 15% | 10% | 7% | 4% | 3% | 6% |
| Date of death of source incompatible with transmission | 3,203 | 57% | 14% | 10% | 6% | 4% | 3% | 6% |
| Migration date of source dates incompatible with transmission | 2,917 | 54% | 15% | 11% | 7% | 4% | 3% | 6% |
| Viral load of source incompatible with transmission | 7,232 | 58% | 12% | 10% | 6% | 3% | 4% | 7% |
| Time elapsed exceeding 16 years | 2,939 | 56% | 14% | 10% | 7% | 4% | 3% | 6% |

Table S1: **Sensitivity analyses on the impact of the additional epidemologic exclusion criteria on transmission flow estimates.**